\documentclass[11pt]{article}
\usepackage{enumitem}
\newlist{examples}{enumerate}{1}
\setlist[examples]{label=(\thetable.\arabic*)}
\usepackage{latexsym}
\usepackage{amssymb}
\usepackage{amsmath}
\numberwithin{equation}{section}
\usepackage{showlabels}
\usepackage{cite}

\RequirePackage{amsthm}
\RequirePackage{amscd}
\RequirePackage{epsfig}
\RequirePackage{graphics}
\RequirePackage{ifthen}
\RequirePackage{varioref}
\usepackage{color}

\newcommand{\beq}{\begin{equation}}
\newcommand{\eeq}{\end{equation}}
\newcommand{\bea}{\begin{eqnarray}}
\newcommand{\eea}{\end{eqnarray}}
\newcommand{\barr}[1]{\begin{array}}
\newcommand{\earr}{\end{array}}
\newtheorem{theorem}{Theorem}[section]
\newtheorem{definition}{Definition}[section]

\newcommand{\bdf}{\begin{definition}}
\newcommand{\edf}{\end{definition}}
\newcommand{\bth}{\begin{theorem}}
\newcommand{\enth}{\end{theorem}}
\newcommand{\pd}{\partial}

\newcounter{abc}
\newenvironment{zoznamrom}{\setcounter{abc}{0}\begin{list}{\roman{abc})}
                     {\usecounter{abc}}}{\end{list}}

\def\c+{\rlap{\ \raisebox{.2ex}{\scriptsize+}}\supset}

\begin{document}
\setcounter{table}{0}

\title{Lie group analysis of a generalized Krichever-Novikov differential-difference equation}
\author{ Decio Levi~$^1$, Eugenio Ricca~$^1$, Zora Thomova~$^3$ and Pavel Winternitz~$^4$}
\maketitle
\noindent
{$^1$ Dipartimento di Matematica e Fisica, Universit\`a degli Studi Roma Tre and Sezione INFN Roma Tre, Via della Vasca Navale, 84, 00146 Roma, Italy \newline e-mail: levi@roma3.infn.it, eugenioricca1@tin.it \\
$^3$ Department of Engineering, Science and Mathematics, SUNY Institute of Technology, 100 Seymour Road, Utica, NY 13502, USA \newline e-mail: thomovz@sunyit.edu\\
$^4$ Centre de recherche math{\'e}matiques and D{\'e}partement de math{\'e}matiques et de statistique, 
Universit{\'e} de Montr{\'e}al, Case postale 6128, succ. centre-ville, Montr{\'e}al, Qu{\'e}bec, H3C 3J7, Canada \newline e-mail: wintern@crm.umontreal.ca}

\begin{abstract}
The symmetry algebra of the differential--difference equation $$\dot u_n = [P(u_n)u_{n+1}u_{n-1} + Q(u_n)(u_{n+1}+u_{n-1})+ R(u_n)]/(u_{n+1}-u_{n-1}),$$ where $P$, $Q$ and $R$ are arbitrary analytic functions is shown to have the dimension $1 \le \mbox{dim}L \le 5$. When $P$, $Q$ and $R$ are specific second order polynomials in $u_n$ (depending on 6 constants) this is the integrable discretization of the Krichever--Novikov equation. We find 3 cases when the arbitrary functions are not polynomials and the symmetry algebra satisfies $\mbox{dim}L=2$. These cases are shown not to be integrable. The symmetry algebras are used to reduce the equations to purely difference ones. The symmetry group is also used to impose periodicity $u_{n+N}=u_n$ and thus to reduce the differential--difference equation to a system of $N$ coupled ordinary three points difference equations. 
\end{abstract}

\section{Introduction}
This article is part of a general program the aim of which is to apply continuous groups to study discrete phenomena. More specifically, the purpose is to use Lie groups to classify, simplify and ultimately solve difference, or differential-difference equations. For recent reviews of this program see \cite{LW2006,Doro-book,W2004,YAM2006,ly2011,k2011,WCUP2011} and for a broader review of the subject of symmetries and integrability of difference equation see the Lecture Notes \cite{lotw}.

One of the applications of Lie groups to differential equations is to perform a symmetry classification. Thus a differential equation, or a family of equations is considered, depending on some unspecified functions $\phi_a(\vec{x}_i, \vec{u}_i,\vec{u}_{\vec{x}}, \cdots)$ where 
$\vec{x} \in \mathbb{R}^p, \vec{u} \in \mathbb{R}^q $ are the independent and dependent variables, respectively. The problem is to determine the Lie point symmetry group of the system for $\phi_a$ generic and then to find all sets of functions $\phi_a$ for which the symmetry group is larger. For examples of such studies treating partial differential equations see e.g. \cite{HorZda2001,CRCHand3,Melesh2006,Cicogna2008,GungorLZ2004,IbTVal1991,
GanTVal2004,HuangZhang2009,bih2011,bih2012,w1992,gung2002,bgl2013,bgo2013}

Such a symmetry classification has been performed for certain discrete dynamical systems (molecular chains) in earlier publications \cite{LW1996,GLW1999,LTW,Zhang2Cong13}. Since these articles were published the theory of symmetries of differential-difference equations has undergone further developments \cite{LWY2002,WCUP2011} and the formalism has been simplified.

The purpose of this article is to perform a symmetry analysis of a specific differential-difference equation which we shall call the generalized Krichever-Novikov differential--difference equation 
\beq
u_{n,t} \equiv \dot{u}_n =\frac{P(u_n)u_{n+1}u_{n-1}+Q(u_n)(u_{n+1}+u_{n-1})+R(u_n)}{u_{n+1}-u_{n-1}}. \label{EYdKN}
\eeq

In (\ref{EYdKN}) the functions $P(u_n), Q(u_n)$ and $R(u_n)$ are arbitrary analytical functions. This equation is of physical and mathematical interest mainly because it generalizes the "Yamilov discretization of the Krichever--Novikov equation" \cite{LWY2011,y83,YAM2006}. In this case $P$, $Q$ and $R$ are restricted to 
\bea \label{1.2}
P_n=\alpha u_n^2 + 2\beta u_n + \gamma, \; Q_n=\beta u_n^2 + \lambda u_n + \delta, \; R_n= \gamma u_n^2 + 2 \delta u_n + \omega, 
\eea
and equation (\ref{EYdKN}) is integrable and allows an infinite number of generalized symmetries. The Lie point symmetries of (\ref{EYdKN}) when $P$, $Q$ and $R$ are arbitrary second order polynomials were studied in \cite{LWY2011}. The dimension of the symmetry algebra $L$ was found to satisfy $2 \le \mbox{dim}L \le 5$. All cases with $\mbox{dim}L=5$ or $4$ are integrable. Among the three cases with $\mbox{dim}L=3$ only two were integrable, among the nine equations with $\mbox{dim}L=2$ exactly three are integrable. In general, integrable equations (\ref{1.2}) tend to have larger Lie point symmetries than non integrable ones. As opposed to non integrable ones, the integrable ones have infinite dimensional algebras of generalized symmetries.

Here we will concentrate on the case when $P$, $Q$ and $R$ are not of the form (\ref{1.2}) or its generalization considered in  \cite{LWY2011} with generic second order polynomials and then compare the results with a symmetry analysis of the "generalized Krichever--Novikov" equation
\beq
v_t=v_{xxx} - \frac{3}{2} \frac{v_{xx}^2}{v_x}+ \frac{f(v)}{v_x}, \label{eq-gKN}
\eeq
where $f(v)$ is an arbitrary analytical function.  A symmetry analysis of (\ref{eq-gKN}) was performed in \cite{BG2010} and \cite{LWY2011}. Equation (\ref{eq-gKN}) is obtained as a continuous limit of (\ref{EYdKN}) as described in \cite{LWY2011}. Thus, we put
\bea
u_n(t)&=&v(x,t)\equiv v, \, \, \, \, \, \, \, x= nh + 6 \frac{t}{h^2}, \, \, \, \, \, \, k=-\frac{12}{h^3} \nonumber \\
P(u_n)&=&k+2h F(v), \, \, \, \, \, \, Q(u_n)=-kv+2h G(v), \, \, \, \, \, \, \, R(u_n)=kv^2+2h H(v) \label{contlim1}
\eea
and  take the limit $h \rightarrow 0$, $n \rightarrow \infty $, $hn$ finite. In the limit we obtain (\ref{eq-gKN}) with
\beq
f(v)=v^2P(v)+2vQ(v)+R(v). \label{fu}
\eeq

The original (integrable) Krichever--Novikov equation \cite{kn80} has the form (\ref{eq-gKN}) where $f$ is a fourth degree polynomial with arbitrary constant coefficients. It firstly appeared in a study of quasi periodic solutions of soliton equations \cite{kn80,kn79}. For further references see \cite{nmpz} and also \cite{LWY2011}.

In Section 2 we derive the determining equations for the symmetries of  (\ref{EYdKN}) and the "allowed transformations" that leave the form of  (\ref{EYdKN}) invariant for arbitrary functions $P$, $Q$ and $R$. The determining equations are solved in Section 3. Eq. (\ref{EYdKN}) is always invariant under time translations and integer shifts in $n$. We find all functions $P$, $Q$ and $R$ for which the symmetry algebra is larger than $X_0=\partial_t$. We present only the cases when $P$, $Q$ and $R$ are not all second order polynomials since such polynomial cases where already treated in \cite{LWY2011}. The 5 non polynomial cases are summed up in Section 4. The symmetry algebra in all non polynomial cases is two--dimensional. Among the 5 cases two are redundant in that they are related by allowed transformations. The 3 representative non polynomial cases are presented in Table 2. In  Section 5 we analyze the results and show that none of the non polynomial cases is integrable. We show how the symmetry group can be used to reduce  (\ref{EYdKN}) to a one variable difference equation. Invariance under shifts of the discrete variable $n$ can be used to impose periodicity, i.e. reduce (\ref{EYdKN}) to a finite system of ordinary differential equations. Some conclusions are presented in the final Section 6.

\section{Determining equations and allowed transformations}

Equation (\ref{EYdKN}) belongs to a class of equations studied in \cite{LWY2002} where it was shown that the vector fields generating their symmetries will have the general form

\beq
X=\tau(t) \pd_t + \phi_n(t, u_n) \pd_{u_n} \label{fieldX}
\eeq
i.e. $\tau(t)$ does not depend on $n$ or $u_n$. The prolongation of $X$ acting on (\ref{EYdKN}) is

\bea
pr X=\tau(t) \pd_t + \sum_{j=n-1}^{n+1} \phi_j(t,u_j) \pd_{u_j} +\phi_n^{(1)} \pd_{\dot{u}_n} \label{prX} \nonumber \\
\phi_n^{(1)} = D_t\phi_n(t,u_n)-\left[ D_t \tau(t) \right] \dot{u}_{n}
\eea
where $D_t$ is the total differentiation operator. We require that $prX$ should annihilate (\ref{EYdKN}) on its solution set. This implies the determining equation

\bea
\phi_{n,t}(u_{n+1}&-&u_{n-1})^2+(\phi_{n,u_n}-\dot{\tau}) \left[ P u_{n+1}u_{n-1}+Q(u_{n+1}+u_{n-1})+R\right](u_{n+1}-u_{n-1}) \nonumber \\
&-&\phi_n \left[ P_{,u_n}u_{n+1}u_{n-1}+Q_{,u_n}(u_{n+1}+u_{n-1})+R_{,u_n} \right](u_{n+1}-u_{n-1}) \nonumber \\
&=&\phi_{n-1} \left[P u^2_{n+1}+ 2 Qu_{n+1}+R \right] -\phi_{n+1} \left[ P u^2_{n-1}+ 2 Qu_{n-1}+R \right.]. \label{DetEqPhi}
\eea

Equation (\ref{EYdKN}) is form invariant under the group of M{\"o}bius transformations of $u_n$ and linear transformations of time
\bea
u_n=\frac{\xi_1 u^*_n+\xi_2}{\xi_3 u^*_n +\xi_4}, \, \, \, \, \, \, t= \theta_1 t^* + \theta_2, \label{Mobius} \\
\xi_1 \xi_4-\xi_2 \xi_3 = \pm 1, \, \, \, \theta_1 \neq 0, \, \, \, \xi_i, \theta_i \in \mathbb{R}.\nonumber
\eea
We shall call (\ref{Mobius}) "allowed transformations".

\noindent Eq. (\ref{EYdKN}) is transformed into
\beq
u^*_{n,t^*} =\frac{\tilde{P}(u^*_n)u^*_{n+1}u^*_{n-1}+\tilde{Q}(u^*_n)(u^*_{n+1}+u^*_{n-1})+\tilde{R}(u^*_n)}{u^*_{n+1}-u^*_{n-1}} \label{U*nt}
\eeq
with
\bea
\tilde{P} (u^*_n)&=&(\xi_3 u^*_n+\xi_4)^2 \left[ P^*(u_n) \xi_1^2
+2 Q^*(u_n)\xi_1 \xi_3 +R^*(u_n) \xi_3^2 \right] \nonumber \\
\tilde{Q}(u^*_n)&= &(\xi_3 u^*_n+\xi_4)^2 \left[ P^*(u_n) \xi_1 \xi_2 +Q^*(u_n)(\xi_1 \xi_4 + \xi_2 \xi_3 ) +R^*(u_n) \xi_3 \xi_4 \right] \label{PQRtrans}\\
\tilde{R}(u^*_n)&=&(\xi_3 u^*_n+\xi_4)^2 \left[ P^*(u_n) \xi_2^2 + 2 Q^*(u_n) \xi_2 \xi_4 +R^*(u_n) \xi_4^2 \right]. \nonumber
\eea
In (\ref{PQRtrans}) $P(u_n), Q(u_n)$ and $R(u_n)$ are to be viewed as the same functions as in (\ref{EYdKN}), however they depend on $u^*_n$ via the M\"{o}bius transformation (\ref{Mobius}) e.g we have 
$P^*(u_n) \equiv P(\frac{\xi_1 u^*_n+\xi_2}{\xi_3 u^*_n +\xi_4})$.

In particular for a pure inversion we have $\xi_2=\xi_3=1, \xi_1=\xi_4=0, \theta_1=1$ and $\theta_2=0$:
\bea \nonumber
\tilde{P}(u^*_n)=(u^*_n)^2R(\frac{1}{u^*_n}), \, \, \, \tilde{Q}(u^*_n)=(u^*_n)^2Q(\frac{1}{u^*_n}), \, \, \, \, \tilde{R}(u^*_n)=(u^*_n)^2 P(\frac{1}{u^*_n})
\eea

From (\ref{PQRtrans}) we see that with no loss of generality we can assume
\beq
P(u_n) \neq 0 \label{Pnonzeo}
\eeq
in (\ref{EYdKN}) (and in (\ref{DetEqPhi})). Indeed if we have $P(u_n)=0$ in (\ref{EYdKN}) then the M\"{o}bius transformation will generate $\tilde{P}(u^*_n) \neq 0$ (unless all three functions satisfy $P(u_n)=Q(u_n)=R(u_n)=0$.

The quantities $u_{n+1}$ and $u_{n-1}$ appear in (\ref{DetEqPhi}) explicitly.  Implicitly  they figure only via $\phi_{n+1}$ and $\phi_{n-1}$, respectively. Taking the derivative $\pd^2_{u_{n+1}} \pd^2_{u_{n-1}}$ we obtain (for $P(u_n) \neq 0$)
\beq
\phi_{n+1,u_{n+1}u_{n+1}}-\phi_{n-1,u_{n-1}u_{n-1}} =0 \label{eq2-9}
\eeq
From (\ref{eq2-9}) and (\ref{DetEqPhi}) we obtain the following result
\begin{theorem}
The Lie algebra of local Lie point symmetries of the Generalized  Krichever--Novikov Differential--Difference equation (\ref{EYdKN}) is realized by vector fields f the form (\ref{fieldX}) with
\bea
\tau&=& \tau_1 t + \tau_0, \, \, \, \qquad \phi_n=\alpha_n+\beta_nu_n+\gamma_n u_n^2  
\label{anbncn} \\
\alpha_n&=&a_1+(-1)^n a_2, \, \, \, \, \, \beta_n=b_1+(-1)^n b_2, \, \, \, \, \, \gamma_n=c_1+(-1)^n c_2 \nonumber
\eea
where $\tau_1, \tau_0, a_1, a_2, b_1, b_2, $ and $ c_1, c_2$ are constants.
\end{theorem}

This theorem was already presented in Ref \cite{LWY2011} for $P, Q$ and $R$ second order polynomials.  Here we have generalized it to arbitrary analytic functions of $u_n$. 

To proceed further we substitute expressions (\ref{anbncn}) into the determining equations (\ref{DetEqPhi}) and collect independent powers of the form 
$u^p_{n+1}u^q_{n-1}$ for $0 \leq p \leq 2, 0 \leq q \leq 2$. Terms $u^2_{n+1}u^2_{n-1}$ cancel, the remaining terms contain an overall factor $(u_{n+1}-u_{n-1})$ that can be dropped. We are left with three equations, the coefficients of $u_{n+1}u_{n-1}, u_{n+1}+u_{n-1}$ and $(u_{n+1})^0 (u_{n-1})^0$, respectively. Since the functions $P(u_n), Q(u_n), R(u_n)$ are assumed to be analytical and $n$-independent, the only dependence on $(-1)^n$ in the determining equations is explicit (i.e. not contained in $P$, $Q$ and $R$). Hence the terms with and without the factor $(-1)^n$ must vanish separately. Finally we obtain six determining equations:
\bea
-a_1 P+c_1 R - (a_1+b_1u_n+c_1u_n^2) Q'+(b_1+2u_nc_1-\tau_1)Q&=&0 \label{e1}\\
a_2P-c_2R-(a_2+b_2u_n+c_2u_n^2)Q'+(b_2+2c_2u_n)Q&=&0 \label{e2}  \\
2c_1Q-(a_1+b_1u_n+c_1u_n^2)P'+(2c_1u_n-\tau_1)P&=&0 \label{e3} \\
-2c_2Q-(a_2+b_2u_n+c_2u_n^2)P'+(2b_2+2c_2u_n)P&=&0 \label{e4} \\
-2a_1Q-(a_1+b_1u_n+c_1u_n^2)R'+(2b_1+2c_1u_n-\tau_1)R&=&0 \label{e5}\\
2a_2Q-(a_2+b_2u_n+c_nu_n^2)R'+2c_2u_nR&=&0 \label{e6}
\eea
We see that the constant $\tau_0$ does not figure in (\ref{e1} -- \ref{e6}) whereas all other constants determining the symmetry vector field
\beq
X=(\tau_1 t+\tau_0) \pd_t+\left[ (a_1+b_1u_n+c_1u_n^2)+(-1)^n (a_2+b_2u_n+c_2u_n^2)\right] \pd_{u_n}
\eeq
are present. Hence the vector field 
\beq
T_0=\pd_t
\eeq
corresponding to $\tau_0=1$ is always present in the symmetry algebra for arbitrary functions $P,Q$ and $R$. The existence of further symmetries imposes conditions on these functions.

Our strategy will be to consider (\ref{e1}--\ref{e6}) as a system of coupled ordinary differential equations for functions $P,Q$ and $R$. We solve them for these functions, treating the constants $\tau_1, a_i, b_i, c_i$ as parameters. Then we use the M\"{o}bius transformations and time dilations to simplify the obtained expressions and re-parametrize them. Finally, we reintroduce the obtained functions into (\ref{e1}--\ref{e6}) and solve these equations for the parameters $\tau_1, a_i, b_i$ and $c_i$ and thus determine the symmetry algebra in each case. We are only interested in cases when at least one of the functions $P, Q$ and $R$ is not a second order polynomial. Once the functions $P$, $Q$ and $R$ and the corresponding symmetry algebras of eq. (\ref{EYdKN}) are obtained we check for redundancies. If different equations (with isomorphic symmetry algebras) can be transformed into each other by an allowed transformation, we keep only one of them in the final representative list.


\section{Solution of determining equation}

In order to solve the determining equations we split the process into several sub-cases. We are interested only in  real solutions ($P$, $Q$, $R \in \mathbb R$).

\subsection{Generic case $c_1 \neq 0$ and $c_2 \neq 0$}
We multiply (\ref{e3}) by $c_2$, (\ref{e4}) by $c_1$ and add the two equations and obtain a first order ODE (depending on 2 parameters $\kappa$ and $\rho$ )
\beq
(u_n^2 +\kappa) P'(u_n) - (2 u_n+ \rho) P(u_n) =0, \, \, \, \, \, \kappa = 1,-1,0, \, \, \, \, \,  \rho \in \mathbb{R} \label{C1-e1}
\eeq
The solutions are 
\begin{eqnarray} \nonumber
\kappa &= 1  \quad &P(u_n)=(u_n^2+1) e^{\rho \arctan u_n} \\ \label{3.2}
\kappa &= -1 \quad &P(u_n)=(u_n-1)^{1+\frac{\rho}{2}} (u_n+1)^{1-\frac{\rho}{2}} \\ \nonumber
\kappa &=0 \quad &P(u_n)= u_n^2 e^{-\frac{\rho}{u_n}}
\end{eqnarray}
We will consider $\rho \neq0 $ first and each value of $\kappa$ separately. Then we consider $\rho=0$ and each $\kappa$ again.
\begin{zoznamrom}
\item
$\rho \neq 0, \kappa = 1$
\bea \nonumber 
P(u_n)&=&(u_n^2+1) e^{\rho \arctan u_n} \\ \label{CI-i}
Q(u_n)&=&(-\frac{1}{2}\rho u_n^2+Au+B)e^{\rho \arctan u_n} \\ \nonumber
R(u_n)&=&\frac{1}{2} \frac{1}{(4+\rho^2)(u_n^2+1)}\left( (\rho^4+6\rho^2+8)u_n^4+Cu_n^3+Du_n^2+Eu_n+F \right) e^{\rho \arctan u_n}.
\eea
Requiring that all equations (\ref{e1}--\ref{e6}) are satisfied implies $c_1=0$,  $\rho=0$ or $\rho$ complex number, i.e. disagree with the assumptions.
\item
$\rho \neq 0, \kappa = -1$ 
\bea \nonumber
P(u_n)&=&(u_n-1)^{1+\frac{\rho}{2}} (u_n+1)^{1-\frac{\rho}{2}} \\ \label{CI-ii}
Q(u_n)&=&\frac{(u_n-1)^{1+\frac{\rho}{2}} (u_n+1)^{1-\frac{\rho}{2}}}{u_n^2-1} \left( -\frac{1}{2}\rho u_n^2+Au+B \right) \\ \nonumber
R(u_n)&=&\frac{(u_n-1)^{1+\frac{\rho}{2}} (u_n+1)^{1-\frac{\rho}{2}}}{2(u_n^2-1)^2(\rho^2-4)}
\left( (\rho^4+6\rho^2+8)u_n^4+Cu_n^3+Du_n^2+Eu_n+F    \right)
\eea
We find again that if we require for all equations (\ref{e1}--\ref{e6}) to be satisfied, there is no solution satisfying all assumptions. 
\item
$\rho \neq 0, \kappa = 0$
\bea
P(u_n)&=&u_n^2 e^{-\frac{\rho}{u_n}} \\
Q(u_n)&=&\left( -\frac{1}{2}\rho u_n^2+Au+B \right) e^{-\frac{\rho}{u_n}} \\
R(u_n)&=&\frac{1}{u_n^2 \rho (\rho^2-4)} \left( C u_n^4+Du_n^3+Eu_n^2+Fu_n+G \right)e^{-\frac{\rho}{u_n}}
\eea
Again equations (\ref{e1}--\ref{e6})  cannot be satisfied. 
To conclude: there is no function $P(u_n) \neq 0$ for which we have $c_1 c_2 \rho \neq 0$.

Considering $\rho =0$ we obtain  three additional cases (for $\kappa = \pm1, 0$). However, in all cases  (\ref{e1}--\ref{e6}) imply that $P(u_n), Q(u_n)$ and $R(u_n)$ are second order polynomials, a case already fully analyzed in \cite{LWY2011}.

%

\end{zoznamrom}


\subsection{Case $c_1 \neq 0$ and $c_2 = 0$}
Equation (\ref{e4}) will be reduced to 
\beq
(b_2u_n +a_2) P'(u_n) -2b_2 P(u_n)=0 \label{eP-case2}
\eeq
\begin{zoznamrom}
\item
$b_2 \neq 0$. \\
This case again leads to $P,Q$ and $R$ that are (at most) second order polynomials.

%
%

\item
$b_2 =0, a_2 \neq 0$ \\
The only solution is $P(u_n)=Q(u_n)=R(u_n)=0$. 

\item
$a_2=0,b_2=0$ \\
We have 3 equations left (\ref{e1}, \ref{e3}, \ref{e5}). Using  allowed transformations we can shift $b_1 \rightarrow 0$ and $c_1 \rightarrow 1$. The solution of the three remaining equations will depend on the value of $a_1$. We need to consider three separate cases $a_1= \pm1, 0$\\
$\bullet  a_1=1$ \\
We have
\bea \nonumber
P(u_n)&=&Ae^{\rho \arctan u} (1+u_n^2), \quad \rho \ne 0,\\ \label{3.15}
Q(u_n)&=&0\\ \nonumber
R(u_n)&=&Ae^{\rho \arctan u} (1+u_n^2)
\eea
We get that $\tau_1=-\rho$ and $a_1=1,b_1=0,c_1=1$ and we already have $a_2=b_2=c_2=0$. \\
$\bullet a_1=-1$ \\
We have
\bea \nonumber
P(u_n)&=&(\frac{1+u_n}{1-u_n})^{\rho} (A+Bu_n+Cu_n^2), \quad \rho \ne 0 \\ \label{3.16}
Q(u_n)&=&-\frac{1}{2} (\frac{1+u_n}{1-u_n})^{\rho}(Bu_n^2+(2A+2C)u_n+B)\\ \nonumber
R(u_n)&=&(\frac{1+u_n}{1-u_n})^{\rho} (C+Bu_n+Au_n^2)
\eea
We also get $\tau_1=2 \rho$ and $a_1=-1,b_1=0,c_1=1$ and we assumed $a_2=b_2=c_2=0$.\\
For special case $\rho=1$ we obtain again polynomial solutions for $P,Q$ and $R$ of (at most) 2nd degree.

$\bullet  a_1=0$ \\
\bea \nonumber
P(u_n)&=&e^{\frac{\rho}{u_n}} (A+Bu_n+Cu_n^2), \quad \rho \ne 0\\ \label{3.17}
Q(u_n)&=&-\frac{1}{2} e^{\frac{\rho}{u_n}} (Bu_n^2+2Au_n)\\ \nonumber
R(u_n)&=&e^{\frac{\rho}{u_n}} Au_n^2
\eea
We obtain $\tau_1=\rho$ and $a_1=0,b_1=0,c_1=1$ and  we assumed $a_2=b_2=c_2=0 $. The coefficient $\rho$ can be rescaled to $\rho=1$ by an allowed transformation $u_n \rightarrow \rho u_n$.
\end{zoznamrom}


\subsection{Case $c_1=0$ and $c_2 \neq 0 $}
Equation (\ref{e3}) reduces to
\beq
(b_1 u_n+a_1) P'(u_n) +\tau_1 P(u_n)=0 \label{c1-0c2-no}
\eeq
and we must substitute its solutions into the system (\ref{e1}--\ref{e6}). Running  through all possible cases we find:
\begin{zoznamrom}
\item 
Unless (\ref{c1-0c2-no}) is solved trivially $P,Q$ and $R$ are second order polynomials.
\item
Eq. (\ref{c1-0c2-no}) is solved trivially leaving $P(u_n)$ arbitrary if $b_1=a_1=\tau_1=0$. We are left with only three equations (\ref{e2}, \ref{e4}, \ref{e6}). Their solution is again a set of second order polynomials. Thus, no new solutions are obtained for $c_2 \ne 0$.

\end{zoznamrom}

\subsection{Case $c_1=0$ and $c_2=0$}
The equations (\ref{e1}--\ref{e6}) simplify to:
\bea
-a_1 P - (a_1+b_1u_n) Q'+(b_1-\tau_1)Q&=&0 \label{c3-e1}\\
a_2P-(a_2+b_2u_n)Q'+b_2Q&=&0 \label{c3-e2}  \\
-(a_1+b_1u_n)P'-\tau_1P&=&0 \label{c3-e3 }\\
-(a_2+b_2u_n)P'+2b_2P&=&0 \label{c3-e4} \\
-2a_1Q-(a_1+b_1u_n)R'+(2b_1-\tau_1)R&=&0 \label{c3-e5}\\
2a_2Q-(a_2+b_2u_n)R'&=&0 \label{c3-e6}
\eea
There are several cases to consider, but only two of them give new results, namely

\begin{zoznamrom}


\item $b_2=0,a_2=0, b_1 \neq 0$ \\
From (\ref{c3-e3 }) we get 
\bea \nonumber
P(u_n)&=&A u_n^{\rho}, \quad \rho \ne 0, \\ \label{3.31}
Q(u_n)&=&B u_n^{1+\rho} \\ \nonumber
R(u_n)&=&C u_n^{2+\rho}.
\eea
We find that $a_1=0$, $ \tau_1 = - \rho b_1$, with $ b_1$ free, and $b_1 \neq 0$.


\item $b_1=0, b_2=0, a_2=0$ \\
We consider $a_1 \neq 0$ as $a_1 =0$  gives us equation $\tau_1 P=0, \tau_1 R=0, \tau_1Q=0$.
For $a_1 \neq 0$ the three remaining equations imply
\bea \nonumber
P(u_n)&=&A e^{\rho u_n}, \quad \rho \ne 0, \\ \label{3.34}
Q(u_n)&=&e^{\rho u_n} (-Au_n+B) \\ \nonumber
R(u_n)&=&e^{\rho u_n} (Au_n^2-2Bu_n+C)
\eea
After substituting back we obtain $\tau_1=-a_1 \rho$, $a_1$ is free and we also had $c_1=c_2=b_1=b_2=a_2=0$. We can put $\rho=1$, $A=1$ and $C=0$ by an allowed transformation of the form $u_n \rightarrow u_n + \beta, \, t \rightarrow \gamma t$.

\end{zoznamrom}


\section{Results of the symmetry classification}

Let us now sum up and analyze the results of Section 3.

\begin{enumerate}
\item We have verified the entire symmetry classification of \cite{LWY2011}, where $P$, $Q$ and $R$ are second order polynomials. A brief summary is given in Table 1.
\end{enumerate}

\vspace{1mm}
\begin{table}[htb]
\centering
\begin{tabular}{|c|c|c|}

\hline 
 {dim $L$} &  {Number of cases} & 
 {Equation in Ref. \cite{LWY2011}} \\ [1ex] 
\hline \hline
5 & 1 & (4.1) \\ 
4 & 2 & (4.2), (4.6) \\ 
3 & 3 & (4.3), (4.4), (4.5) \\
2 & 9 & (4.8) (7 cases), (4.10), (4.11) \\
[0.75ex] \hline 
\end{tabular}
\caption{Summary of cases with $P$, $Q$ and $R$ polynomials (of second order) and symmetry algebra $L$ satisfying dim$L \ge 2$.}
\label{t1}
\end{table}

\begin{enumerate}
\addtocounter{enumi}{1}
\item We have obtained 5 cases when $P(u_n)$ is not a second order polynomial and the symmetry algebra satisfies dim$L \ge 2$. They correspond to (\ref{3.15}--\ref{3.17}, \ref{3.31}) and (\ref{3.34}), respectively. All of them have two dimensional non--Abelian symmetry algebras of the form
\bea \label{4.1}
X_0=\partial_t, \qquad X_1=t \partial_t +(A u_n^2+B u_n + C) \partial_{u_n},
\eea 
where $A$, $B$ and $C$ are some specific constants. The vector fields ($u_n^2 \partial_{u_n}, \,  u_n \partial_{u_n}, \,  \partial_{u_n}$) span an $sl(2,\mathbb R)$ algebra. The M\"obius transformations (\ref{Mobius}) represents the Lie group $SL(2,\mathbb R)$ that acts on these vector fields by inner automorphisms. Hence we can use this M\"obius transformation to transform the term $(A u_n^2 + B u_n +C)\partial_{u_n}$ into a representative of one of the three distinct one--dimensional subalgebras  of $sl(2, \mathbb R)$, namely: $(1+u_n^2) \partial_{u_n}$ (rotation $o(2)$),  $u_n \partial_{u_n}$ (pseudorotations $o(1,1)$), or $\partial_{u_n}$ (unipotent transformations). This suggests that some of the five obtained generalized Krichever--Novikov equations may be redundant, i.e. related to each other by M\"obius transformations and this is indeed the case. 

The compact subalgebra $o(2)$ occurs just once, namely in the case (\ref{3.15}). The functions $P$, $Q$ and $R$   depend on one constant $A$. By a time dilation we can set $A=1$ and have
\bea \label{4.2}
X_1 = t \partial_t - \frac{1}{\rho} (1+u_n^2) \partial_{u_n}.
\eea
The non compact subalgebra $o(1,1)$  occurs twice. First in (\ref{3.16}) with
\bea \label{4.3}
X_1=t \partial_t +\frac{1}{2\rho} (-1 + u_n^2) \partial_{u_n},
\eea
and secondly in (\ref{3.31}) with
\bea \label{4.4}
X_1=t\partial_t -\frac{1}{\rho} u_n \partial_{u_n}.
\eea
The M\"obius transformation $u_n \rightarrow \frac{1-u_n}{1+u_n}$ takes the vector field (\ref{4.3}) and also the functions $P$, $Q$ and $R$ in  (\ref{3.16}) into those in (\ref{3.31}) (with a redefinition of the constants).

The nilpotent subalgebra also occurs twice, in (\ref{3.17}) with 
\bea \label{4.5}
X_1=t \partial_t +  u_n^2 \partial_{u_n},
\eea
and in (\ref{3.34}) with
\bea \label{4.6}
X_1=t\partial_t -  \partial_{u_n}.
\eea
The M\"obius transformation $u_n \rightarrow \frac{1}{u_n}$ takes the vector field (\ref{4.6}) into (\ref{4.5}) and also the functions $P$, $Q$ and $R$ of (\ref{3.17}) into those of (\ref{3.34}).

Finally we are left with three inequivalent cases, summed up in Table 2. In each case  we can calculate the continuous limit as indicated in (\ref{contlim1}, \ref{fu}) of the Introduction. The results are included in Table 2. Using a rescaling of time we can always rescale $A \rightarrow 1$.

\end{enumerate}

\vspace{1mm}
\begin{table}
\centering

\begin{tabular}{|c|c|c|c|}
\hline 

 {$N_0$} &  {The equation} & 
 {Basis elements} & Continuous limit \\ [1ex] 
\hline \hline
1 & (\ref{5.1}) & $X_1=t\pd_t-\frac{1}{\rho} (1+u_n^2) \pd_{u_n}$ & $(v^2+1)^2e^{p \arctan v}$ \\ 
2 & (\ref{5.2}) & $X_1= t \pd_t - \frac{1}{\rho} u_n \pd_{u_n}$ & $ v^p$ \\ 
3 & (\ref{5.3}) & $X_1= t\partial_t - \partial_{u_n}$&  $ e^v$\\
[0.75ex] \hline 
\end{tabular}
\caption{Representative list of equations with non polynomial functions $P$, $Q$ and $R$ and dim$L >1$ (in all cases we have dim$L=2$). The last column shows the function $f(v)$ in the continuous limit (\ref{eq-gKN}).}
\label{t2}
\end{table}

\section{Analysis of non polynomial cases}
The generalized Krichever--Novikov differential--difference equation is always invariant under time translations (since the coefficients do not depend on $t$). The main result of this article is that we have identified all cases, when the symmetry group is larger, i.e. its Lie point symmetry algebra $L$ contains at least one additional elements. In addition to known cases \cite{LWY2011} summed up in Table 1 we have identified three new ones given in Table 2.

Let us now analyze the new differential--difference equations, namely
\bea \label{5.1}
\dot u_n &=& e^{\rho \arctan u_n}\frac{(u_{n+1} u_{n-1} +1)(1+u_n^2)}{u_{n+1}-u_{n-1}}, \qquad \rho \ne 0, \\ \label{5.2}
\dot u_n &=& u_n^{\rho} \frac{u_{n+1}u_{n-1} + Au_n(u_{n+1}+u_{n-1})+Bu_n^2} {u_{n+1}-u_{n-1}}, \qquad \rho \ne 0, \\ \label{5.3}
 \dot u_n &=& e^{u_n}\frac{u_{n+1}u_{n-1} + (-u_n+A)(u_{n+1}+u_{n-1})+u_n^2 - 2 A u_n}{u_{n+1}-u_{n-1}}.
\eea
 \begin{enumerate}
\item The first question concerns their integrability. In the original Yamilov discretization of the Krichever--Novikov equation \cite{msy,y83,YAM2006} it was  shown that for $P$, $Q$ and $R$ satisfying (\ref{1.2}) equation  (\ref{EYdKN}) is integrable and allows an infinity of commuting generalized symmetries. Let us now consider  (\ref{5.1}), (\ref{5.2}) and (\ref{5.3}) from this point of view.
 From \cite{lw97} we know that any equation of the Volterra type $\dot u_n =f(u_{n-1}, u_n, u_{n+1})$ which has higher generalized symmetries has to satisfy a set of conditions given in (3.10) of \cite{lw97}. The first of them is 
\bea
\frac{\delta}{\delta u_n} \partial_t \log \frac{\partial f}{\partial u_{n+1}}=0, \label{var-con}
\eea
where the variational derivative operator is given by $\frac{\delta}{\delta u_n}a_n(u_{n-i},\cdots, u_{n+j})=\sum_{k=n-i}^{n+j} \frac{\partial a_k}{\partial u_n}$ with $i,j$ positive numbers. Applying this condition to the different  $f$'s corresponding to the three non polynomial cases (\ref{5.1} -- \ref{5.3}) we find that this condition is never satisfied and consequently the three non polynomial cases are not integrable. This was to be expected as they are not contained in the complete classification of Volterra type equations up to point and Miura transformations performed by Yamilov \cite{y83} and reviewed in \cite{YAM2006}.
\item The symmetries of the obtained differential--difference equations can be used to perform symmetry reduction.

Eq. (\ref{5.1}) is invariant under the transformations induced by the vector field  $X=-\rho t \partial_t +$\newline$+(u_n^2+1)\partial_{u_n}$. Invariant solutions will have the form
\bea \label{5.5}
u_n =  \tan (\gamma_n - \frac{1}{\rho} \log t).
\eea
Substituting (\ref{5.5}) into (\ref{5.1}) we find that $\gamma_n$ must satisfy the nonlinear recursion relation
\bea \label{5.6}
\tan(\gamma_{n+1}-\gamma_{n-1}) = - \rho e^{\rho \gamma_n}.
\eea

From eq. (\ref{5.2}) the vector field $X=-\rho t \partial_t + u_n \partial_{u_n}$ provides the reduction formula
\bea \label{5.7}
u_n = \gamma_n t^{-\frac{1}{\rho}},
\eea
where $\gamma_n$ must satisfy the recursion relation
\bea \label{5.8}
\frac{\gamma_n^{\rho -1}}{\gamma_{n+1}-\gamma_{n-1}} \left [ \gamma_{n+1} \gamma_{n-1} + A \gamma_n (\gamma_{n+1} + \gamma_{n-1}) + B \gamma_n \right ] = -\frac{1}{\rho}.
\eea

Finally, for eq. (\ref{5.3}) the vector field $X=t\partial_t - \partial_{u_n}$ provide the reduction formula 
\bea \label{5.9}
u_n=\gamma_n - \log t,
\eea
and the following recursion relation for $\gamma_n$:
\bea \label{5.10}
\frac{e^{\gamma_n}}{\gamma_{n+1}-\gamma_{n-1}} \left [ \gamma_{n+1} \gamma_{n-1} +(A - \gamma_n)(\gamma_{n+1} + \gamma_{n-1}) + \gamma_n^2 - 2 A \gamma_n \right ] = -1.
\eea
Thus, in each case invariance under the subgroup corresponding to the vector field $X_1$ leads to the reduction of the differential--difference equation to a three term nonlinear difference equation.

It is interesting to compare the symmetry algebra of eq. (\ref{5.1}), (\ref{5.2}) and (\ref{5.3}) with those of their continuous limits, given in Table 1 of Ref. \cite{LWY2011}. The symmetry algebras $\{ X_0, X_1 \}$ of Table 2 of this article survive (and remain the same) in this limit. An additional vector field $X_2=\partial_x$ appears in the limit. The corresponding group of transformation $n \rightarrow n+N, \, N \in \mathbb Z$ is also present in the discrete case (\ref{EYdKN}) (for arbitrary functions $P$, $Q$ and $R$). It is this symmetry that allows the construction of "periodic systems" satisfying 
\bea \label{5.11}
u_{n+N}(t)=u_n(t).
\eea
For $N=1$ and $N=2$  (\ref{EYdKN})  implies $\dot u_n=0$, and we obtain  trivial difference equations.
\newline For $N\ge 3$ we obtain a coupled system of $N$ differential equations. For example for $N=3$ the system is
\bea \nonumber
\dot u_0 &=& \frac{P(u_0) u_1 u_2 +Q(u_0)(u_1+u_2)+R(u_0)}{u_2-u_1},\\ \label{5.12}
\dot u_1 &=& \frac{P(u_1) u_2 u_0 +Q(u_1)(u_2+u_0)+R(u_1)}{u_0-u_2},\\ \nonumber
\dot u_2 &=& \frac{P(u_2) u_0 u_1 +Q(u_2)(u_0+u_1)+R(u_2)}{u_1-u_0}.
\eea

\section{Conclusions}
The main result of this paper are the three equations (\ref{5.1}--\ref{5.3}) and their symmetry algebras $\{X_0=\partial_t, \, X_1 \}$ with $X_1$ given in (\ref{4.2}), (\ref{4.4}) or (\ref{4.6}), respectively. As shown in Section 5, these symmetry algebras can always be put to good use, just as in the case of differential equations.

Let us end with a rather general comment. Eq.  (\ref{EYdKN}), studied in this article, is a particular discretization (from two continuous variables to one continuous and one discrete) of the generalized Krichever--Novikov equation. The form of the discretization was inspired by Yamilov's  discretization \cite{y83} of the original integrable Krichever--Novikov equation \cite{kn80}.  Yamilov's discretization preserved integrability, in particular an infinite Abelian algebra of generalized symmetries. Here we have shown that this discretization also preserves all point symmetries on a regular (equally spaced) lattice. This is to be contrasted with the general fact that a complete discretization of an ODE or PDE preserving point symmetries requires the introduction of non uniform symmetry adapted lattices \cite{Doro-book,lotw,W2004,LW2006}.

An alternative point of view is that the PDE (\ref{eq-gKN}) is the continuous limit of the differential--difference equation (\ref{EYdKN}) and that all point symmetries of the continuous limit  are obtained  from the point  symmetries of (\ref{EYdKN}). This is not the case for all differential--difference equations. Indeed it was shown in \cite{hlrw} that the Toda lattice has the potential Korteweg--de Vries equation as a continuous limit. Two of the five  Lie point symmetries of this KdV equation are however obtained from generalized symmetries of the Toda lattice, not from point ones.

The relation between point symmetries of difference equations or differential--difference equations and their continuous limits needs a more general study.  
\end{enumerate}

\section*{Acknowledgement}
D.L. has been partly supported by the Italian Ministry of Education and Research, 2010 PRIN ÒContinuous and discrete nonlinear integrable evolutions: from water waves to symplectic mapsÓ. D.L., E.R. and Z.T. thank the  CRM for hospitality. The research of P.W. is partly supported by a grant from NSERC of Canada. 



\begin{thebibliography}{99}

\bibitem{bgl2013}P. Basarab--Horwath, F. G\"ung\"or, V. Lahno, Symmetry classification of third-order nonlinear evolution equations. Part I: Semi-simple algebras, {\it Acta Appl. Math.} {\bf 24} (2013) 1--48.

\bibitem{bgo2013}P. Basarab--Horwath, F. G\"ung\"or, C. \"Ozemir, Infinite--dimensional symmetries of a general class of variable coefficient evolution equations in 2+ 1 dimensions, {\it J.  Phys.: Conference Series} {\bf 474} (2013) 012010.

\bibitem{HorZda2001}P. Basarab--Horwath, V. Lahno and R. Zhdanov, The Structure of Lie Algebras and the Classification Problem for Partial Differential Equations, {\it Acta Appl. Math.} {\bf 69} (2001)  43--94.

\bibitem{bih2012}A. Bihlo, E. Dos Santos Cardoso-Bihlo and R. O. Popovych, Complete group classification of a class of nonlinear wave equations, {\it J. Math. Phys.} {\bf 53} (2012)123515.

\bibitem{bih2011}A. Bihlo and R. O. Popovych, Lie symmetry analysis and exact solutions of the quasigeostrophic two-layer problem, {\it J. Math. Phys.} {\bf 52} (2011) 033103.  

\bibitem{BG2010} M.S. Bruzon and M.L. Gandarias, Classical and nonclassical reductions for the Krichever-Novikov equation, in  { \it ICNAAM, AIP Conf. Proc.}, {\bf 1281} (2010) 2147--2150.

\bibitem {Cicogna2008} G. Cicogna, Symmetry classification of quasi-linear PDE's containing arbitrary functions, {\it Nonlinear Dynam.} {\bf 51} (2008) 309--316.

\bibitem{Doro-book} V.A. Dorodnitsyn,  {\it Applications of Lie Groups to Difference Equations}, CRC Press, Boca Raton,  2011.

\bibitem{k2011}V.A. Dorodnitsyn and R. Kozlov, Lagrangian and Hamiltonian formalism for discrete equations: symmetries and first integrals,   in {\it Symmetries and Integrability of Difference Equations}, LMS Lecture Series, eds. D. Levi, P.J. Olver, Z. Thomova, P. Winternitz, CUP, Cambridge, 2011,  7--49.

\bibitem{GanTVal2004} M.L. Gandarias, M. Torrisi and  A. Valenti, Symmetry classification and optimal systems of a non-linear wave equation, {\it Int. J. Non-Linear Mech.} {\bf 39} (2004) 389--398.

\bibitem{w1992}J. P. Gazeau and P. Winternitz, Symmetries of variable coefficient KortewegÐde Vries equations, {\it J. Math. Phys.} {\bf 33} (1992) 4087--4102 . 

\bibitem{GLW1999} D. Gomez-Ullate, S. Lafortune and P. Winternitz,  Symmetries of discrete dynamical systems involving two species. {\it J. Math. Phys.}, {\bf 40} (1999) 2782--2804.

\bibitem{GungorLZ2004} F. G{\"u}ng{\"o}r, V. I. Lahno and R. Z. Zhdanov, Symmetry classification of KdV-type nonlinear evolution equations, {\it J. Math. Phys.} {\bf 45} (2004)  2280--2313.
 
\bibitem{gung2002}F. G\"ung\"or and P. Winternitz, Generalized KadomtsevÐPetviashvili equation with an infinite-dimensional symmetry algebra, {\it J. Math. Anal.  Appl.} {\bf 276} (2002) 314--328.

\bibitem{hlrw}R. Hernandez-Heredero, D. Levi, M. Rodriguez, and P. Winternitz. Lie algebra contractions and symmetries of the Toda hierarchy . {\it J. Phys. A Math. Gen.} {\bf 33} (2000) 5025--5040.

\bibitem{HuangZhang2009} D. Huang and H. Zhang, Preliminary group classification of quasilinear third-order evolution equations, {\it Appl. Math. Mech.} {\bf 30} (2009)  275--292.

\bibitem{CRCHand3} N.H. Ibragimov editor, {\it Handbook of Lie Group Analysis of Differential Equations. Vol.2: Applications in Engineering and Physical Sciences}, CRC Press, Boca Raton, 1996.

\bibitem{IbTVal1991} N.H. Ibragimov, M. Torrisi and A. Valenti, Preliminary group classification of equations $v_{tt}= f (x, v_x) v_{xx}+ g (x, v_x)$, {\it J. Math. Phys.} {\bf 32} (1991) 2988--2995.

\bibitem{kn79}I.M. Krichever and S.P. Novikov, Holomorphic bundles and non linear equations. Finite zone solutions of rank 2. {\it Dokl. Akad. Nauk SSSR} {\bf 247} (1979), 33--36, in English: {\it Sov. Math. Dokl.} {\bf 20} (1979) 650--651.

\bibitem{kn80}I.M. Krichever and S.P. Novikov,  Holomorphic Bundles over Algebraic Curves, and Nonlinear Equations,  {\it Russ. Math. Surv.} {\bf 35} (1980) 53--80. English translation of {\it Uspekhi Mat. Nauk} {\bf 35} (1980) 47--68.

\bibitem{LTW} S. Lafortune, S. Tremblay and P. Winternitz,  Symmetry
classication of diatomic molecular chains. {\it J. Math. Phys.}, {\bf 42} (2001)5341--5357.

\bibitem{lotw} D. Levi, P.J. Olver, Z. Thomova and P. Winternitz eds., {\it Symmetries and Integrability of Difference Equations}, LMS Lecture Series, CUP, Cambridge, 2011.

\bibitem{LW1996} D. Levi and P. Winternitz,  Symmetries of discrete dynamical systems. {\it J. Math. Phys.} {\bf 37} (1996) 5551--5576.

\bibitem{LW2006} D. Levi and P. Winternitz,  Continuous symmetries of difference equations. {\it J. Phys. A  Math. Theor.}  {\bf 39} (2006) R1--R63.

\bibitem{LWY2002} D. Levi, P. Winternitz and R.I. Yamilov,  Lie point symmetries of differential-difference equations. {\it J. Phys. A Math.Theor.} {\bf 43} (2010)    292002 (14 pp). 

\bibitem {LWY2011} D. Levi, P. Winternitz  and R.I. Yamilov,  Symmetries of the continuous and discrete Krichever-Novikov equation. {\it SIGMA}, {\bf 7}, (2011) 097  (21 pp).

\bibitem{lw97}D. Levi and R.I. Yamilov, Conditions for the existence of higher symmetries of evolutionary equations on the lattice, {\it J. Math. Phys.} {\bf 38} (1997) 6648--6674.

\bibitem{ly2011}D. Levi and R. I. Yamilov, Generalized Lie symmetries for difference equations,  in {\it Symmetries and Integrability of Difference Equations}, LMS Lecture Series, eds.D. Levi, P.J. Olver, Z. Thomova, P. Winternitz, CUP, Cambridge 2011, 160--190.
 
\bibitem{Melesh2006} S.V. Meleshko, {\it Methods for Constructing Exact Solutions of Partial Differential Equations: Mathematical and Analytical Techniques with Applications to Engineering}, Springer, Berlin, 2006.

\bibitem{msy}A.V. Mikhailov, A.B. Shabat and R.I. Yamilov, The symmetry approach to the classification of nonlinear equations. Complete lists of integrable systems {\it Uspekhi Mat. Nauk} {\bf 42} (1987) 3--53;
English transl.:  {\it Russian Math. Surveys} {\bf 42} (1987) 1--63.


\bibitem{W2004} P. Winternitz, Symmetries of discrete systems. In {\it Discrete Integrable Systems}, B. Grammaticos,
Y. Kosmann-Schwarzbach, and T. Tamizhmani, editors,  Lecture Notes in Physics 644, Springer, Berlin, 2004,   185--243.

\bibitem{WCUP2011}P. Winternitz, Symmetry preserving discretization of differential equations and Lie point symmetries of differential-difference equations, in {\it Symmetries and Integrability of Difference Equations}, LMS Lecture Series, eds. D. Levi, P.J. Olver, Z. Thomova, P. Winternitz, CUP, Cambridge,  2011, 292--341.

\bibitem{y83}R.I. Yamilov, Classification of discrete evolution equations, {\it Uspekhi Mat. Nauk} {\bf 38} (1983) 155-156.

\bibitem{YAM2006} R.I. Yamilov,  Symmetries as integrability criteria for differential difference equations {\it J. Phys. A Math. Gen.} {\bf 39} (2006) R541--R623.

\bibitem{nmpz}V.E. Zakharov, S.V. Manakov, S.P. Novikov, and L.P. Pitaevskii, {\it Teoriya solitonov. Metod obratnoi zadachi}, Nauka, Moscow 1980, in English: {\it The theory of solitons. The method of the inverse problem}, Plenum Press, New York, 1984.

\bibitem{Zhang2Cong13} X. Zhang, Y. Cong and H. Zhang, Preliminary group classification of the nonlinear differential-difference equations
{\it J. Math. Anal. Appl.}, {\bf 399}, (2013)  638--649.





\end{thebibliography}
\end{document}